\documentclass[journal]{IEEEtran}
\usepackage{amsmath,amsfonts,graphicx,tabularx,color}
\usepackage{amssymb,textcomp,mathtools,amsthm}
\usepackage{bm,upgreek,algorithm,hyperref,algorithmic}
\usepackage{multirow,booktabs,hhline,array}
\usepackage{cite,url,makecell,setspace,listings}
\RequirePackage{doi}

\definecolor{codegreen}{rgb}{0,0.6,0}
\definecolor{codegray}{rgb}{0.5,0.5,0.5}
\definecolor{codepurple}{rgb}{0.58,0,0.82}
\definecolor{backcolour}{rgb}{0.95,0.95,0.92}

\lstdefinestyle{mystyle}{
    backgroundcolor=\color{backcolour},   
    commentstyle=\color{codegreen},
    keywordstyle=\color{magenta},
    numberstyle=\tiny\color{codegray},
    stringstyle=\color{codepurple},
    basicstyle=\ttfamily\scriptsize,
    breakatwhitespace=false,         
    breaklines=true,                 
    captionpos=b,                    
    keepspaces=true,                 
    numbers=left,                    
    numbersep=5pt,                  
    showspaces=false,                
    showstringspaces=false,
    showtabs=false,                  
    tabsize=2
}

\theoremstyle{definition}

\theoremstyle{plain}

\theoremstyle{remark}

\title{FRA-RIR: Fast Random Approximation of the Image-source Method}

\author{Yi~Luo, Jianwei~Yu}

\begin{document}

\maketitle
\setlength{\abovedisplayskip}{2pt}
\setlength{\belowdisplayskip}{2pt}
\setlength{\abovedisplayshortskip}{2pt}
\setlength{\belowdisplayshortskip}{2pt}

\begin{abstract}
The training of modern speech processing systems often requires a large amount of simulated room impulse response (RIR) data in order to allow the systems to generalize well in real-world, reverberant environments. However, simulating realistic RIR data typically requires accurate physical modeling, and the acceleration of such simulation process typically requires certain computational platforms such as a graphics processing unit (GPU). In this paper, we propose FRA-RIR, a fast 
random approximation method of the widely-used image-source method (ISM), to efficiently generate realistic RIR data without specific computational devices. FRA-RIR replaces the physical simulation in the standard ISM by a series of random approximations, which significantly speeds up the simulation process and enables its application in on-the-fly data generation pipelines. Experiments show that FRA-RIR can not only be significantly faster than other existing ISM-based RIR simulation tools on standard computational platforms, but also improves the performance of speech denoising systems evaluated on real-world RIR when trained with simulated RIR. A Python implementation of FRA-RIR is available online\footnote{\url{https://github.com/yluo42/FRA-RIR}}.
\end{abstract}


\section{Introduction}
\label{sec:intro}
The simulation of room impulse response (RIR) filter plays an important role in the training of various modern speech processing systems. Systems trained without reverberant data can hardly generalize well to real-world scenarios \cite{kinoshita2013reverb}, and a good RIR simulator can further improve the performance by augmenting the anechoic data with various RIR filters \cite{kim2017generation}.

A wide range of systems rely on an \textit{offline training} configuration, where a fixed number of training and development samples are generated in advance and kept unchanged during the training phase. For the generation of simulated reverberant speech samples, each anechoic utterance requires a simulated RIR filter to create a training sample. With the rapid growth of the scale of the data, such generation process becomes time and storage consuming and further limits the amount of available training data that can be used by the systems. As a consequence, \textit{online training} with \textit{on-the-fly} data simulation becomes important as it can not only generate infinite training data but also requires no extra storage.

The fast and efficient simulation of realistic RIR filters, however, remains challenging. One of the most widely-used methods is the \textit{image-source method (ISM)} \cite{allen1979image}, where the propagation and the reflection of the sound sources are calculated by virtual sound images generated by mirroring the original sound sources by the room boundaries (e.g., floors and walls). However, such simulation typically assumes an empty rectangular or parallelepiped room and a fixed absorption rate of all boundaries, which may not be able to simulate realistic RIR filters that matches the room conditions in the real world where different furniture and materials may result in complicated reflection patterns. Moreover, the calculation of the sound paths can be complex and time consuming. A standard method to accelerate ISM is to use GPU-accelerated implementations \cite{fu2016gpu, diaz2020gpurir}, where the physical modeling part can benefit from specific computational platforms. Moreover, to improve the quality of the RIR filters, diffuse-based methods were proposed to better model late reverberation \cite{lehmann2009diffuse, tang2020improving}, and ray-tracing-based methods were explored to use explicit room modeling to calculate the sound paths \cite{kulowski1985algorithmic, funkhouser2004beam, scheibler2018pyroomacoustics}. Neural networks, especially generative adversarial networks (GANs) \cite{goodfellow2014generative}, were also adopted to refine simulated RIR filters to approximate the distributions of the real-recorded RIR filters \cite{ratnarajah2020ir, ratnarajah2021ts, ratnarajah2022fast}. Although many of these methods have proven effective in certain applications and platforms, their usage in on-the-fly data simulation have not been fully evaluated and may still limited by their speed, complicity and the quality of the generated RIR filters.

In this paper, we propose a simple method to approximate the physical modeling of the sound propagation and reflection process in ISM, which we refer it to as \textit{Fast Random Approximation of RIR (FRA-RIR)}. FRA-RIR is particularly designed for the data simulation process in on-the-fly training configuration, which aims at the on-the-fly generation of realistic RIR filters without the requirement of any specific computational devices or platforms. Instead of explicit calculating the virtual sound paths, FRA-RIR randomly approximates the paths as well as the their reflection patterns to generate an energy-rescaled dirac comb at a higher sample rate, and then downsamples it to the target sample rate to generate the actual RIR filter. The relationship between the sound propagation distance and the reflection is determined via heuristic assumptions and evaluated by grid search on the hyperparameters. With a standard desktop-level CPU, FRA-RIR can generate more realistic RIR filters than existing ISM-based method up to 130 times faster, which enables fully on-the-fly data simulation. Moreover, when only trained with simulated RIR filters, speech enhancement models trained with FRA-RIR can also achieve on par or better performance than other ISM-based RIR simulation methods.

The rest of the paper is organized as follows. Section~\ref{sec:FRA-RIR} introduces the proposed FRA-RIR method and provides corresponding visualization and comparison with other existing RIR simulation methods. Section~\ref{sec:config} describes the experiment configurations. Section~\ref{sec:result} presents the results on speech denoising and dereverberation task. Section~\ref{sec:config} concludes the paper.

\section{Fast Random Approximation of the Image-source Method}
\label{sec:FRA-RIR}
\subsection{Image-source Method Recap}
\label{sec:ISM}
We adopt the definition of an RIR filter in \cite{kim2017generation}:
\begin{align}
    h[n] = \frac{1}{d_0} \delta \left[n - \left\lceil \frac{d_0f_s}{c_0} \right\rceil \right] + \sum_{i=1}^I \frac{r^{g_i}}{d_i} \delta \left[n - \left\lceil \frac{d_if_s}{c_0} \right\rceil \right]
\label{eqn:image}
\end{align}
where $I$ denotes the total number of virtual sound sources, $d_0$ denotes the distance of the direct-path sound source, $d_i$ denotes the distance from the $i$-th virtual sound image to the receiver, $r$ denotes the reflection coefficient of the surface, $g_i$ denotes the number of the reflections of the $i$-th sound source, $f_s$ denotes the target sample rate, and $c_0$ denotes the sound velocity. We follow the same estimation of the reflection coefficient via the Eyring's empirical equation \cite{beranek2006analysis, kim2017generation}:
\begin{align}
    r = \sqrt{1 - \left(1 - e^{-0.16R/ T_{60}}\right)^2}
\label{eqn:reflect}
\end{align}
where $R$ denotes the ratio between the volume and the total surface area of the room, and $T_{60}$ denotes the reverberation time that takes for the sound to decay by 60 dB in the room.

To ensure a sufficiently high temporal resolution on the time difference of arrival (TDOA) of different virtual sound sources, $h[n]$ should be generated in a sufficiently high sample rate. Given that the target sample rate of the RIR filter is $f_s$, $h[n]$ should be generated at sample rate $r_hf_s$, where $r_h>1$ is the rescaling factor. Following the configuration in \cite{kim2017generation}, $h[n]$ is first downsampled to an intermediate sample rate $r_lf_s$ with $1<r_l<r_h$ being another rescaling factor, and then a high-pass filter with a cut-off frequency of 80 Hz is applied to remove the unwanted low-frequency components \cite{allen1979image, kim2017generation}. The filtered RIR filter is then downsampled again to sample rate $f_s$ to serve as the final output to be convolved with the actual sound source.

\subsection{Fast Random Approximation of RIR}
\label{sec:FRA}

FRA-RIR bypasses the explicit calculation of equation~\ref{eqn:image} by \textit{sound path sampling}. Compared to the standard ISM method, FRA-RIR makes three core modifications:
\begin{enumerate}
    \item We randomly sample the room-related statistics $R$ instead of calculating it via the length, width and the height of an empty room.
    \item We replace the explicit calculation of $d_i$ by sampling it from a probability distribution.
    \item We replace the explicit calculation of $g_i$ by defining it as a function of $d_i$ with random perturbations.
\end{enumerate}

\subsubsection{Simulating Room-related Statistics}
\label{sec:room}

The the ratio between the volume and the total surface area of the room $R$, which we define as the room-related statistics, is typically calculated based on the length, width and height of the room. It also implicitly assumes an empty room so that the calculation of the total surface area only considers the walls. To enable the approximation of a realistic room-related statistics, we first randomly sample a $T_{60}$ within range $[0.1, 0.8]$, and then we directly sample $R$ within range $[0.1, 1.2]$ instead of explicitly calculating its value. We set the upperbound of $R$ to be $1.2$ based on the assumption that a larger room leads to a higher upperbound for $R$, and the value $1.2$ is calculated from an ideal empty rectangular room with length, width and height of 12~m, 12~m and 4~m, respectively. The reflection coefficient is then calculated by equation~\ref{eqn:reflect}.

\subsubsection{Distance Simulation in FRA-RIR}
\label{sec:dist}

For an empty rectangular or parallelepiped room, the distance between a virtual sound source and the receiver can be directly computed via their 3D coordinates. However, when there are extra surfaces in the room, the reflection of the sound sources can be highly complicated, and the coordinates of the images of the original sound source with respect to all the available surfaces can be extremely hard to accurately calculate. FRA-RIR randomly samples the distance ratio $DR_i \triangleq d_i/d_0$ between the $i$-th virtual and direct-path sound source following the probability distribution defined by a simple quadratic function:
\begin{align}
    P(x) = \begin{dcases*}
            \frac{3x^2}{\beta^3 - \alpha^3}, & $\alpha \leq x \leq \beta$ \\
            0, & otherwise
            \end{dcases*}
\end{align} 
where $0\leq\alpha\leq\beta\leq1$ are scalars controlling the range of the distribution. Intuitively, using a quadratic function to generate the probability distribution ensures that the number of distant virtual sound sources increases as their distance $d_i$ increases. Note that the quadratic function can be replaced by other functions and we simply select it due to its simplicity. $\hat{DR}_i \in [\alpha, \beta]$ is first sampled from $P(x)$, and $DR_i$ is generated by linearly rescaling $\hat{DR}_i$ to range $[1, c_0T_{60}/d_0]$, where $c_0T_{60}$ is the maximum distance for a virtual sound source to travel with the given sound velocity and reverberation time:
\begin{align}
    DR_i = 1 + \frac{\alpha}{\beta-\alpha}(\frac{\hat{DR}_i}{\alpha} - 1)(\frac{c_0T_{60}}{d_0}-1)
\end{align}
We empirically set $\alpha=0.2$ and $\beta=1$ in our configuration due to its effectiveness in our experiments. The actual travelling distance $d_i$ can then be calculated by $d_i=DR_i\cdot d_0$, and we uniformly sample $d_0$ within range $[0.2, 12]$ m.

\subsubsection{Reflection Simulation in FRA-RIR}
\label{sec:reflect}

Given the reverberation time $T_{60}$, direct-path distance $d_0$, the reverberation coefficient $r$ and the sound velocity $c_0$, we first calculate the maximum number of reflections a virtual sound source may have to decay by 60 dB through reflection:
\begin{align}
    RR_{max} = (\text{log}_{10}\,c_0T_{60} - \text{log}_{10}\,d_0 - 3) /\text{log}_{10}\,r
\end{align}

We then sample the number of reflections $g_i \in [1, RR_{max}]$ by defining it as a function of $d_i$, and further add a random perturbation to it:
\begin{align}
\begin{split}
    p_i &\sim \mathcal{U}(a, b) \\
    g_i &= 1 + (\frac{d_i}{c_0T_{60}})^2 \cdot (RR_{max} - 1) + p_i \cdot d_i^{\tau} \\
    g_i &= \text{max}(\text{min}(g_i, RR_{max}), 1)
\end{split}
\end{align}
where $\mathcal{U}$ denotes the uniform distribution, $p_i$ denotes the random perturbation on the number of reflections, and $\tau>0$ denotes the distance shrinkage factor. The simulation of $g_i$ is based on the heuristic assumption that images with longer propagation distances may encounter more reflections, and images with a similar overall propagation distances may also have different numbers of reflections. We empirically set $a=-2$, $b=2$ and $\tau=0.2$.

\subsubsection{Generation of the RIR Filter}
\label{sec:RIR}

The generation of $h[n]$ is straightforward after the sampling of $d_i$ and $g_i$. We first initialize the RIR filter $h$ to an all-zero vector of length $L\triangleq\lceil T_{60}r_hf_s\rceil$, and then add each of the virtual sources to the filter:
\begin{align}
    q_i &= \text{min}(\lceil \frac{d_i}{c_0}r_hf_s \rceil, L-1) \\
    h[q_i] &= h[q_i] + \frac{r^{g_i}}{d_i}
\end{align}
We set $g_i=0$ for $i=0$ (i.e., the direct-path sound source). In tasks where the system is required to perform dereverberation, an early-reverberation-RIR filter is needed to serve as the target for the early reverberation component. We define the context of $[-6, 50]$ ms around the direct-path sound source as the early reverberation component:
\begin{align}
    h_e[n] = \begin{dcases*}
             h[n], & $ - \lceil\frac{6r_hf_s}{1000} \rceil \leq n - \lceil \frac{d_0}{c_0}r_hf_s \rceil \leq \lceil\frac{50r_hf_s}{1000} \rceil $ \\
             0, & otherwise
             \end{dcases*}
\end{align}
$h[n]$ and $h_e[n]$ are then passed to the same downsampling--highpass filtering--downsampling process as in \cite{kim2017generation}. We set $r_h=64$ and $r_l=8$ in our configuration to follow the configuration in \cite{kim2017generation}.

\begin{figure}[!hbt]
	\small
	\centering
	\includegraphics[width=\columnwidth]{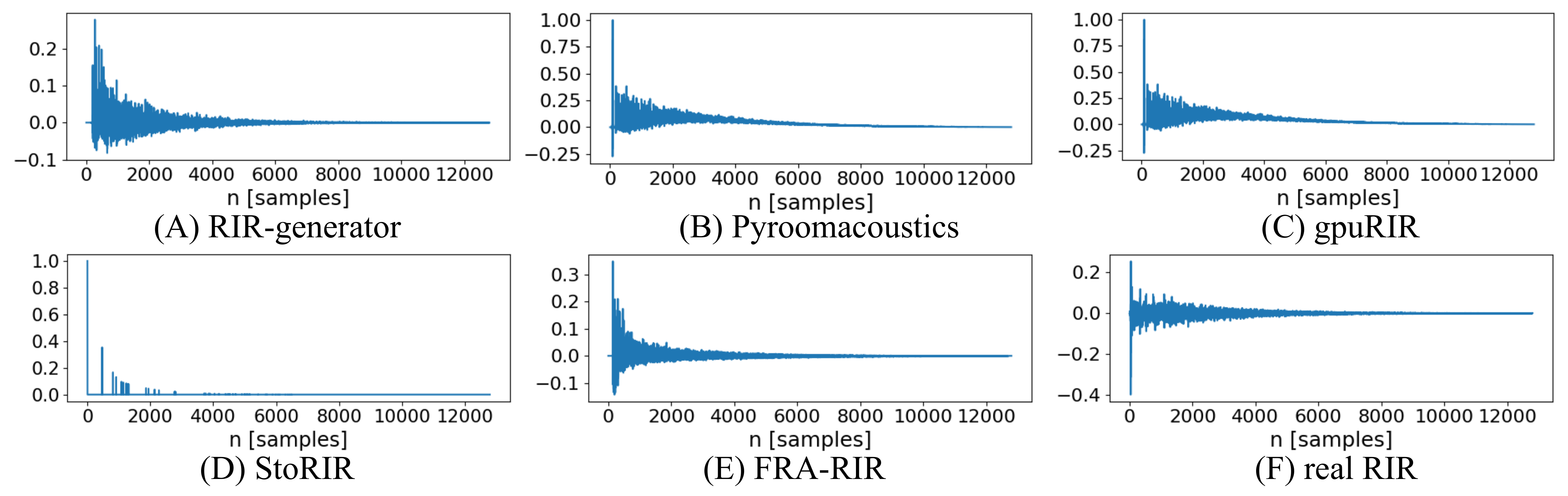}
    \hfill
    \includegraphics[width=\columnwidth]{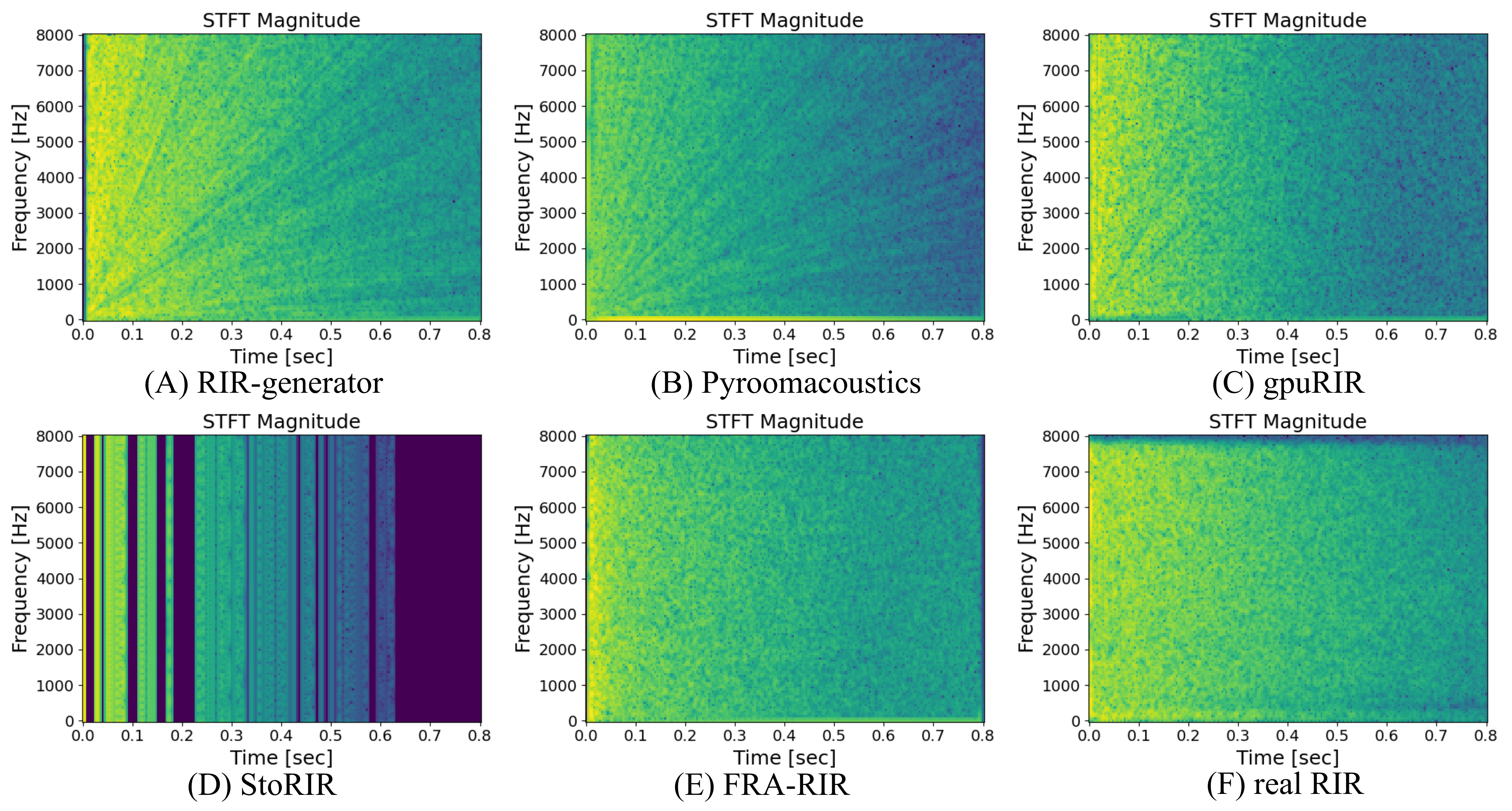}
\caption{Visualization of the simulated and real RIR filters by their waveforms and magnitude spectrograms.}
\label{fig:RIR}
\end{figure}

\vspace{-0.2cm}
\subsection{Visualization}
\label{sec:visualization}

We provide visualizations of multiple RIR filters generated by the proposed FRA-RIR method and compare them to RIR filters generated by other RIR simulation tools at 16k Hz sample rate:
\begin{enumerate}
    \item RIR-generator (RIR-gen) \cite{rir_gen}: RIR-generator is one of the most widely-used implementation of ISM. We use the default configuration provided in the official Python implementation\footnote{\url{https://github.com/audiolabs/rir-generator}}.
    \item Pyroomacoustics (PRA) \cite{scheibler2018pyroomacoustics}: The ISM-based room simulation module in Pyroomacoustics assumes shoebox rooms and considers walls as perfect reflectors. We use the default configuration provided in the official documentation\footnote{\url{https://pyroomacoustics.readthedocs.io/en/pypi-release/pyroomacoustics.room.html}}. We do not use consider the hybrid ISM and ray tracing method in this paper as recommended in the toolbox.
    \item gpuRIR \cite{diaz2020gpurir}: gpuRIR is a GPU-accelerated tool with additional functionalities such as diffuse late reverberation modeling, negative reflection coefficients and fractional delay. We use the default configuration provided in the official example\footnote{\url{https://github.com/DavidDiazGuerra/gpuRIR}}.
    \item StoRIR \cite{masztalski2020storir}: StoRIR uses a random energy-rescaled impulse train to estimate the RIR filter. Although it is not an ISM-based method, we select it as one of the comparable methods as it also generates the RIR filters in a stochastic way. We use the default configuration provided in the official implementation\footnote{\url{https://github.com/SRPOL-AUI/storir}}.
\end{enumerate}
We also randomly selected a real RIR in the BUT ReverbDB dataset\footnote{VUT\_FIT\_D105/MicID01/SpkID05\_20170901\_S/04} \cite{szoke2019building} and use it to compare with the simulation outputs. Figure~\ref{fig:RIR} shows the waveforms and the frequency responses evaluated by the magnitude spectrograms of the simulated and real RIR filters calculated with a window size of 256 point and hop size of 64 point, respectively. We can observe that compared to RIR-gen and PRA, the gpuRIR method which makes use of diffused late reverberation simulation can generate more realistic RIR filters, however we can still observe irregular patterns at the transition between early and late reflections. Due to the characteristic of the filter generated by StoRIR, its frequency pattern diverges from the realistic RIR the most among all the filters. FRA-RIR generates a more realistic frequency pattern in both early and late reverberations compared to the real RIR.

\section{Experiment configurations}
\label{sec:config}
\begin{table*}[!htb]
    \centering
    \small
    \caption{Comparison of different RIR simulation tools with offline (\textit{Off.}) and online (\textit{On.}) training configurations on speech denoising and dereverberation tasks. * corresponds to the configuration where GPU is required.}
    \begin{tabular}{c|cc|cc|cc|cc|cc|cc|cc|c}
    \toprule
        \multirow{3}{*}{Method} & \multicolumn{6}{c|}{Denoising}  & \multicolumn{6}{c|}{Denoising \& Dereverberation}& \multirow{3}{*}{Speed (s)} \\
        \cline{2-13}
        & \multicolumn{2}{c|}{SNR (dB)} & \multicolumn{2}{c|}{PESQ} & \multicolumn{2}{c|}{STOI} & \multicolumn{2}{c|}{SNR (dB)} & \multicolumn{2}{c|}{PESQ} & \multicolumn{2}{c|}{STOI} \\
        \cline{2-13}
        & Off. & On. & Off. & On.   & Off. & On.  & Off. & On. & Off. & On. & Off. & On. & \\
        \hline
        Mixture & \multicolumn{2}{c|}{-2.1}  & \multicolumn{2}{c|}{1.83} & \multicolumn{2}{c|}{65.5} & \multicolumn{2}{c|}{-6.7} & \multicolumn{2}{c|}{1.57} & \multicolumn{2}{c|}{62.5} & -- \\ 
        \hline
        gpuRIR \cite{diaz2020gpurir} & 8.1 & -- & 2.18 & -- &75.0 & --& 2.2 & -- & 1.77 & -- & 68.7 & -- & 0.02* \\
        RIR-gen \cite{rir_gen} & 8.1  & -- & 2.17  & -- & 74.8 & -- & 2.4 & -- & 1.76 & -- & 68.7 & -- & 9.4 \\
        PRA \cite{scheibler2018pyroomacoustics} & 7.7 & 8.6  & 2.09 &  2.21 & 74.0 & 75.8 & 2.2 & 2.7 & 1.68 & 1.81 & 67.3 & 68.1 & 0.88 \\
        StoRIR \cite{masztalski2020storir} & 8.0 & 8.6 & 2.03 & 2.23 & 73.8 & 75.4  & \textbf{2.5} &  2.7 & 1.79 & 1.87 & 69.9 & 70.1 & 0.89 \\
        \hline
        FRA-RIR & \textbf{8.3}  & \textbf{8.7}  & \textbf{2.22}  & \textbf{2.31}  & \textbf{75.4} & \textbf{76.0} & \textbf{2.5} & \textbf{2.9} & \textbf{1.85} & \textbf{1.95} & \textbf{71.0} & \textbf{71.5} & 0.08 \\
    \bottomrule
    \end{tabular}
    \label{tab:result-enhance}
\end{table*}

\subsection{Data Configuration}
\label{sec:data}
We evaluate the effectiveness of the proposed FRA-RIR method in the speech denoising and joint speech denoising and dereverberation tasks. We perform both offline data simulation and on-the-fly data simulation with the aforementioned five RIR simulation methods. During the training phase, the simulated RIR filters are convolved with randomly sampled speech utterances from AISHELL-2 \cite{du2018aishell} and DNS challenge \cite{reddy2021interspeech} at a sample rate of 16k Hz, and we truncate the utterances to 6 seconds. One or two noise utterances are also randomly sampled from the DEMAND \cite{thiemann2013diverse}, MUSAN \cite{snyder2015musan} and DNS challenge datasets, and the RIR filters are simulated and convolved accordingly. For other ISM-based methods, the room size is randomly sampled from $3\times3\times3\ m^3$ to $12\times12\times4\ m^3$ (length$\times$width$\times$height). All noise utterances are summed to generate a single noise signal, and the signal-to-noise ratio (SNR) between the speech and noise signals is randomly sampled between [-8, 6] dB. The number of utterances in the offline training dataset is 50000 ($\approx$110 hours). During the test phase, real RIR filters from the DNS challenge dataset is used to simulate 500 utterances. All other configurations are kept identical for all RIR simulation methods.

\subsection{Model Configuration}
\label{sec:model}

We use the CLDNN architecture for all experiments \cite{sainath2015convolutional}. We use 4 convolution layers, 2 LSTM layers and 1 output layer in the model, and we generate complex ratio mask (cRM) \cite{williamson2015complex} to extract the speech signals. Interested readers may refer to the original paper for the details of the model architecture. We use 32~ms window size, 16~ms hop size and Hanning window for STFT. All models contain 3.3M parameters.

\subsection{Training and Evaluation Configurations}

The training objective for all models is the combination of a waveform-level L1 loss and a spectrogram-level L1 loss on both real and imaginary parts. We use the Adam optimizer \cite{kingma2014adam} with the initial learning rate of 0.001, and we decay the learning rate by a factor of 0.5 if no best training model is found in 3 consecutive epochs. We set the maximum number of training epochs to be 50 and the training will be early stopped when no best validation model is found in 5 consecutive epochs. All of our experiments are conducted on one single server with 8 NVIDIA Tesla P40 GPUs and 64 CPU cores using Pytorch \cite{paszke2019pytorch} with a per-GPU batch size of 16. For online training with on-the-fly data simulation, the RIR filters are simulated in parallel using CPU with 8 workers per data loader. We set the number of effective utterances the same in offline and online configurations for a fair comparison.

For evaluation, we report ({\color{black}{SNR}}), perceptual evaluation of speech quality (PESQ)\footnote{https://github.com/vBaiCai/python-pesq} \cite{rix2001perceptual} and short-time objective intelligibility (STOI)\footnote{https://github.com/mpariente/pystoi} \cite{taal2010short} to measure the speech denoising and dereverberation performance of models trained with different RIR simulation methods.

\section{Results and analysis}
\label{sec:result}
Table~\ref{tab:result-enhance} summarizes the performance of models trained with data generated by different RIR simulation methods. We can first observe that for the offline training configuration, models trained with all RIR simulation methods have similar denoising and dereverberation performance, while FRA-RIR is slightly better than the others. For the online training configuration with on-the-fly data simulation, all models have improved performance compared with the offline training configuration, while FRA-RIR is still slightly better than the others on all three evaluation metrics. Moreover, the RIR simulation speed for FRA-RIR is significantly faster than all other methods except for gpuRIR which requires a GPU to perform the simulation\footnote{We did not perform on-the-fly simulation for gpuRIR and RIR-gen, because gpuRIR does not support CPU-only simulation in the data loaders and RIR-gen is too slow to finish the training procedure.}. Since FRA-RIR not only saves the storage by on-the-fly simulation but also significantly accelerated the training of the model and generalizes well in realistic RIR filters, it proves the potential of FRA-RIR as an effective data augmentation method for the training of speech processing systems.

\section{Conclusion}
\label{sec:conclusion}
In this paper, we proposed the fast random approximation of room impulse response (FRA-RIR), a fast RIR simulation method based on the image-source method (ISM) for data augmentation purpose in the training of speech processing systems. FRA-RIR bypassed the explicit calculation of the virtual sound sources in ISM by randomly sampling the virtual sound source distances and their reflection patterns. Without the need of a specific computational device or platform, FRA-RIR can generate RIR filter up to 110 times faster than existing ISM-based RIR simulation methods on a desktop-level CPU, enabling on-the-fly data simulation for training various speech processing models. Experiment results on speech denoising and jointly denoising and dereverberation tasks showed that models trained with FRA-RIR can achieve on par or better performance than other RIR simulation tools with a significantly faster simulation speed. Future works include the application and validation of the effectiveness of FRA-RIR in other types of speech and audio processing tasks.

\bibliographystyle{IEEEbib}
\bibliography{refs}

\end{document}